\documentclass[aps,prd,endfloats*]{revtex4}
\usepackage{graphicx,amssymb,amsfonts}
\newcommand{\be}{\begin{equation}}
\newcommand{\ee}{\end{equation}}
\newcommand{\ba}{\begin{eqnarray}}
\newcommand{\ea}{\end{eqnarray}}

\newcommand{\pnu}{{p_\nu}}
\newcommand{\pnub}{{p_{\bar\nu}}}
\newcommand{\pf}{{p_f}}
\newcommand{\enu}{{E_\nu}}
\newcommand{\enub}{{E_{\bar\nu}}}

\begin{document}
\title{Neutrino Counter Nuclear Weapon}
\author{Alfred Tang}
\email{atang@fnal.gov}
\date{\today}

\begin{abstract}
Radiations produced by neutrino-antineutrino annihilation at the $Z^0$ pole
can be used to heat up the primary stage of a thermonuclear warhead and
can in principle detonate the device remotely.  Neutrino-antineutrino
annihilation can also be used as a tactical assault weapon to target hideouts
that are unreachable by conventional means. 
\end{abstract}
\maketitle

\section{Introduction}
Nuclear weapon is the most destructive kind among weapons of mass destruction.
Hiroshima and Nagasaki are lessons in history that shall never be repeated.
Since the end of World War II, world leaders had tried to control the
proliferation of nuclear weapons by
political means such as the Nuclear Non-proliferation Treaty
in 1968.  Many countries did not sign the treaty.  In fact it seems that
more and more countries are pursuing nuclear weapon programs nowadays.
After September 11, the concern is that nuclear weapons will fall into
the hands of terrorists.  Strategically speaking the importance of a counter
nuclear weapon may soon rival that of the nuclear weapon itself.  The
purpose of this paper is to explore the possibility of a neutrino counter
nuclear weapon technology.  The idea of using neutrinos to detonate or melt
a nuclear weapon was first proposed by H. Sugawara, H. Hagura and
T. Sanami~\cite{sugawara}.  Their futuristic design is based on a 1~PeV
neutrino beam operating at 50~GW.  It is unlikely that such an intense ultra
high energy neutrino beam can be realized in the near future.  Even if such a
neutrino beam is made available, its radiation hazard
will render it politically nonviable.
Other proposals such as installing neutron detectors at the border to intercept
nuclear materials had been considered.  The current trend of non-proliferation
policy is focused on monitoring the production of fissile fuels.  Research
is being conducted to use anti-neutrino detectors to this end~\cite{suekane}.
Anti-neutrinos are produced in nuclear fission through beta decay.  They
are indicators of the fissile fuel composition of the nuclear reactor.
Neutrino signatures of the fissile fuels cannot be tampered with by virtue of
the very small reaction cross section of neutrinos at low energy.  On the
other hand, the small reaction probability also means small detection
probability so that large detectors are needed to detect them.
A sample idea is to deploy hundreds of kilo-ton liquid scintillor
detectors at 1000~km distance from the reactor to monitor the reactor
anti-neutrino spectrum.  The challenges of using anti-neutrino to monitor
reactor are that (1) a rogue nation will not voluntarily allow IAEA to build
anti-neutrino detectors around its reactors, (2) the number of anti-neutrino
detectors must increase 4 folds for every doubling of reactor-detector distance,
and (3) reactors are not needed if a rogue nation opts for uranium instead of
plutonium bombs.  For these reasons, anti-neutrino detectors are probably
not the ultimate solution to non-proliferation.  Another possible
non-proliferation strategy is to
develop a technology that counters nuclear weapons.

This paper proposes an alternative idea for a neutrino counter nuclear
weapon that shares some similarities with the idea presented in
Reference~\cite{sugawara} but is technologically
feasible, relatively cheap and safe.  The present idea is to focus a neutrino
beam and an antineutrino beam together in a small region to allow them to
annihilate so that high energy radiations are released as reaction products.
The radiations cause neutron spallation in the sub-critical nuclear
material and initiate fission reactions.  The plutonium heats up,
ignites the chemical explosive around the fissile (fissionable material)
in the primary stage of a
thermonuclear warhead and subsequently detonates the nuclear weapon.
The reason of thinking about neutrino for this
application is that neutrino cannot be shielded.  It can hit a target such as
a nuclear submarine from the other side of the globe and can penetrate a deep
underground concrete bunker and missile silo.
Since neutrino can penetrate the planet to reach a nuclear
weapon on the other side of the globe near the speed of light, a neutrino
counter nuclear weapon is in principle untraceable and indefensible.  
It is suggested that a neutrino counter nuclear weapon is 100\%
effective~\cite{bing}.

The trade-off of developing a
counter weapon is the introduction of a new weapon.  If the new weapon is
less destructive than the original weapon, an ethical argument can be made
in support of its development.  If remote detonation of a nuclear weapon
is made possible by a neutrino counter weapon, a nuclear weapon in the
homeland becomes a liability so that there is a real strategic incentive
to reduce the stockpile.  In that case, there will be a much more
convincing political reason to promote non-proliferation.  This work aims to
study the theoretical feasibility of the neutrino counter nuclear weapon as
a first step in this direction.  The use of neutrino as a tactical assault
weapon will also be discussed.

\section{Theoretical Preparation}
\begin{figure}
\includegraphics[scale=0.6]{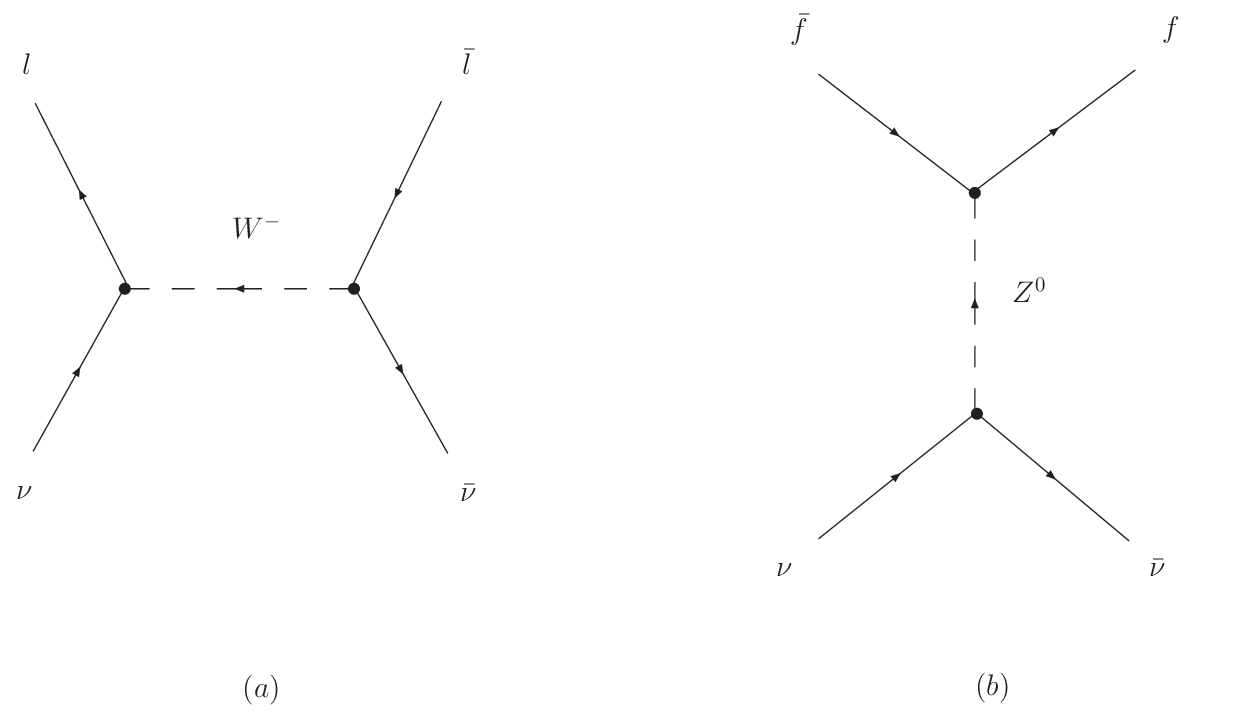}
\caption{\label{fd}
Tree-level Feynman diagrams of neutrino-antineutrino $(\nu\bar\nu)$
annihilation with
(a) a charged current and (b) a neutral current.  The products of $\nu\bar\nu$
annihilation is a lepton-antilepton pair $(l\bar{l})$ in the charge current
case and a fermion-antifermion pair $(f\bar{f})$ in the neutral current case.
The fermion $f$ can be a lepton, neutrino or quark.  In the case of quark
production, mesons will be created by hadronization.}
\end{figure}

One of the properties of neutrino is that it has a vanishingly small
interaction cross section so that it is the most penetrative radiation known
to be in existence today.  The invariant amplitude $\mathcal M$ is proportional
to the weak boson (massive spin-1) propagator~\cite{griffiths,collins,weinberg}
\be
{-i(g_{\mu\nu}-q_\mu q_\nu/M^2)\over q^2-M^2+i\epsilon},
\label{prop}
\ee
where $q$ is the transfer momentum, $M$ is the weak boson mass and the
infinitesimal $\epsilon$ moves the pole away from the branch so that
Eq.~(\ref{prop})
is mathematically well defined.  The weak bosons are $W^\pm$ and $Z^0$.
The corresponding masses are $M_W=80.403\pm0.029$~GeV and
$M_Z=91.1876\pm0.0021$~GeV~\cite{pdg}.  In the case of an unstable particle
with a total decay width $\Gamma$, it is customary to make the replacement
$M^2\to M^2+iM\Gamma$ according to the Breit-Wigner resonance formula
so that Eq.~(\ref{prop}) becomes~\cite{peskin}
\be
{-i(g_{\mu\nu}-q_\mu q_\nu/M^2)\over q^2-M^2-iM\Gamma}.
\label{prop2}
\ee
It is said that the propagator in Eq.~(\ref{prop2}) violates gauge invariance
and that slightly more elaborate modifications are
needed~\cite{kielanowski,nowakowski}.
It is also suggested that Eq.~(\ref{prop2}) is not theoretically justified
in quantum mechanics and quantum field theory and is motivated mostly by
phenomenology~\cite{leo}.  Eq.~(\ref{prop2}) also violates Lorentz invariance
in a subtle way.  The decay width $\Gamma$ is typically
measured in the rest frame of the particle and is related to the lifetime of
the particle $\tau_0$ in its rest frame as $\tau_0=1/\Gamma$.
Lifetime is not Lorentz invariant and can be dilated ($\tau=\gamma\tau_0$).
Traditionally $\Gamma$ in Eq.~(\ref{prop2}) is taken to be a constant just
like $M$.  As it will be shown later, the width term can be eliminated
from the final result by special relativistic considerations.  For the sake
of simplicity, only Eq.~(\ref{prop}) with $\epsilon=0$ will be used in
the calculations of the invariant amplitude $\mathcal M$ in this section.
When $q^2\ll M^2$, the propagator in Eq.~(\ref{prop})
is reduced to $ig_{\mu\nu}/M^2$ and is weighted down by a very
heavy weak boson mass such that the invariant amplitude becomes very
small.  It is the reason why neutrino is so non-interactive even though the
weak coupling constants $g_W$ and $g_Z$ are both larger than the
electromagnetic coupling constant $g_e$~\cite{griffiths},
\ba
g_W&=&{g_e\over\sin\theta_w},\\
g_Z&=&{g_e\over\sin\theta_w\cos\theta_w},\\
G_F&=&{\sqrt{2}\over8}\,\left({g_W\over M_W}\right)^2,
\ea
where $g_e=7.297352568\times10^{-3}$, $\theta_w$ is the Weinberg angle
$(\sin^2\theta_w=0.23122)$ and $G_F$ is the Fermi constant
($G_F=1.16637\times10^{-5}$~GeV$^{-2}$)~\cite{pdg}.
As a corollary, large $M_W$ is the reason why the inverse
beta decay cross section is vanishingly small.  The present work focuses
only on the tree-level diagrams such as $\nu\bar\nu\to l\bar l$ and
$\nu\bar\nu\to f\bar f$ ($l$ for ``lepton'', $f$ for ``fermion'')
as shown in Fig.~\ref{fd}.  Reactions such as $\nu\bar\nu\to\gamma\gamma$
is possible but is a higher-order diagram~\cite{crewther}.
It turns out that
a $W^\pm$ pole is impossible for a charged current in the Feynman diagram of
Figure~\ref{fd}a because the transfer momentum is always null
$(q^2=0)$ or spacelike $(q^2<0)$
as it will be shown later.  In the case of neutral current in
Figure~\ref{fd}b, the transfer momentum can be timelike $(q^2>0)$
so that a $Z^0$ pole is possible at $q^2=M^2_Z$.  The $Z^0$ pole
has been experimentally observed in the spectrum of $e^+e^-$ scattering 
at LEP and SLAC since 1989~\cite{weinberg} and in
$p\bar p$ scattering experiments at Fermilab~\cite{cdf}.
This work takes advantage of the $Z^0$ pole to maximize the
neutrino-antineutrino annihilation cross section to produce high energy
radiation in a maximally efficient way.

\begin{figure}
\includegraphics[scale=0.6]{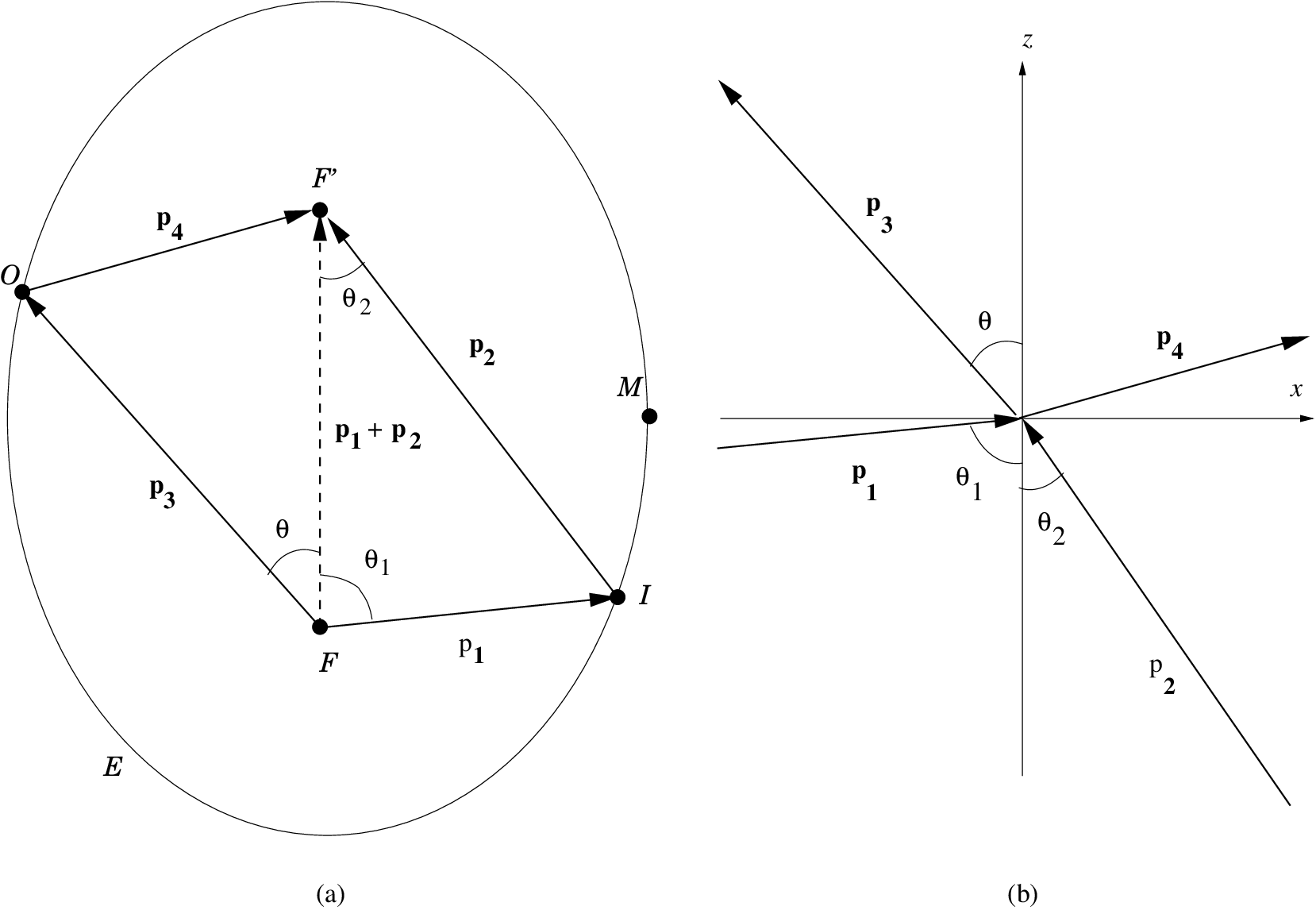}
\caption{\label{ellipse}
Graphical representation of incoming 3-momenta $\mathbf{p_1}$ and
$\mathbf{p_2}$ and outgoing 3-momenta $\mathbf{p_3}$ and $\mathbf{p_4}$.
The point $\mathcal O$ lies on the ellipsoid $\mathcal E$ in part (a) of the
figure.  The point $\mathcal I$ is fixed by the incoming 3-momenta.
Part (a) shows a special case in which all the 3-momenta lie on the same plain.
$\mathcal F$ and $\mathcal{F'}$ are the foci of the ellipsoid.  The point
$\mathcal M$ lies on the minor axis of $\mathcal E$.  If
${\mathcal M}={\mathcal I}$, $\theta_1=\theta_2$.
Momentum is conserved on $\mathcal E$, namely $p_1+p_2=p_3+p_4$.
Part (b) illustrates the Cartesian coordinates used to define the
set of 3-momenta.}
\end{figure}

For the charged current in Figure~\ref{fd}a, The flavor of the outgoing lepton
$l$ (antilepton $\bar l$) must match that of the incoming $\nu$ ($\bar\nu$) at
each vertex.  Mixed flavors are possible in a
reaction as a whole.  For example, $\nu_e \bar\nu_e \to e^- e^+$,
$\nu_e \bar\nu_\mu \to e^- \mu^+$ and all other combinations are allowed.
The invariant amplitude of $\nu\bar\nu \to l\bar l$ in Fig.~\ref{fd}a is
\ba
{\mathcal M}_W&=&-{g_W^2\over8(q^2-M_W^2)}\left[\bar{u}_\nu\gamma^\mu
(1-\gamma^5)u_l\right]\left(g_{\mu\nu}-{q_\mu q_\nu\over M_W^2}\right)
\left[\bar{v}_{\bar l}\gamma^\nu(1-\gamma^5)v_{\bar\nu}\right]\nonumber\\
&\simeq&-{g_W^2\over8(q^2-M_W^2)}\left[\bar{u}_\nu\gamma^\mu
(1-\gamma^5)u_l\right]
\left[\bar{v}_{\bar l}\gamma_\mu(1-\gamma^5)v_{\bar\nu}\right].
\label{invw}
\ea
The transfer momentum is $q=p_l-\pnu=\pnub-p_{\bar l}$.  The
approximations made in the second step of Eq.~(\ref{invw}) are the small masses
of neutrino $(m_\nu\sim0)$ and lepton $(m_l\ll M_W)$.  The invariant
amplitude is squared, summed over final spins and averaged over initial spins.
This calculation is simplified by applying the usual Casimir's
trick and trace theorems.
Neutrino has helicity $h=-1$ but antineutrino has $h=+1$ so that there is only
1 spin state for each initial neutrino or antineutrino.  The Casimir
trick sums over all initial and final spin states by default.  By
``averaging over initial spins'',
it simply means that the sum must be divided by the proper
factor to avoid over-counting.  This way the sum over initial
spins must be divided by 2 for each initial neutrino or antineutrino.
For a reaction involving 2 incoming neutrinos, the proper invariant amplitude
square $\langle{\mathcal M}^2\rangle$ needs to be divided by 4.  With the
standard procedure outlined above, Eq.~(\ref{invw})
can be squared, summed and averaged to give
\be
\langle{\mathcal M}^2_W\rangle=\left[{g^2_W\over q^2-M_W^2}\right]^2
(p_l\cdot\pnub)(\pnu\cdot p_{\bar l}).
\label{invw2}
\ee
Since the present calculation is focused on high energy $(q^2\to M^2)$, the
rest mass of neutrino and lepton can be neglected ($m_\nu\to0$ and
$m_l\to0$).  In the lab frame, the 4-momenta can be parametrized as
$p=(E,\,p_x,\,p_y,\,p_z)$ in such a way that 4-momentum is conserved by
definition:
\ba
p_1&\simeq&E_1(1,\,\sin\theta_1,\,0,\,\cos\theta_1),\label{p1}\\
p_2&\simeq&E_2(1,\,-\sin\theta_2,\,0,\,\cos\theta_2),\\
p_3&\simeq&E_3(1,\,\sin\theta\cos\phi,\,\sin\theta\sin\phi,\,\cos\theta),
\label{p3}\\
p_4&\simeq&p_1+p_2-p_3.\label{p4}
\ea
The $z$-axis in Eqs.~(\ref{p1})--(\ref{p4}) is chosen to point in the
direction of the sum of the initial 3-momenta ${\mathbf p}_1+{\mathbf p}_2$.
The outgoing 3-momenta ${\mathbf p}_3$ and ${\mathbf p}_4$ can rigidly rotate
around the $z$-axis through the azimuthal angle $\phi$ and still conserve
3-momentum.  For the sake of illustration, Fig.~\ref{ellipse} shows a special
case in that all of the incoming and outgoing 3-momenta lie on the same
plain.  The distance between the foci $\mathcal F$ and $\mathcal{F'}$
corresponds to the
vectorial sum ${\mathbf p}_1+{\mathbf p}_2={\mathbf p}_3+{\mathbf p}_4$ so
that 3-momentum is conserved automatically.  Assuming the high energy
condition $E_i\gg m_i$,
$E_i\simeq|{\mathbf p}_i|$ for $i\in\{1,\,2,\,3,\,4\}$.  The ellipsoid
$\mathcal E$ constrains the point $\mathcal O$ in such a way that
$|{\mathbf p}_1|+|{\mathbf p}_2|=|{\mathbf p}_3|+|{\mathbf p}_4|$
so that energy is conserved by $E_1+E_2=E_3+E_4$.  In essence,
the construction in Fig.~\ref{ellipse} guarantees the conservation
of 4-momentum as long as the high energy condition is met.
In the special case of $\theta_1\to0$ and $\theta_2\to0$,
the approximations $E_i\simeq|{\mathbf p}_i|$
made in Eqs.~(\ref{p1})--(\ref{p4}) may become invalid
and should be replaced by the exact parametrizations
\be
p_i=(E_i,\,\mathbf{|p|}_i\sin\theta_i\cos\phi_i,\,
\mathbf{|p|}_i\sin\theta_i\sin\phi_i,\,\mathbf{|p|}_i\cos\theta_i).
\label{psmall}
\ee
However small values of $\theta_1$ and $\theta_2$ do not have any practical
advantage and will not be used in this work.
For the charged current diagram of Fig.~\ref{fd}a, $p_1=\pnu$, $p_2=\pnub$,
$p_3=p_l$ and $p_4=p_{\bar l}$.

The azimuthal angle $\phi$ is not constrained by kinematics and is
completely random.  The transfer
momentum square $q^2$ in the charged current diagram of Fig.~\ref{fd}a can be
expressed in terms of the parametrizations in Eqs.~(\ref{p1})--(\ref{p4}) as
\be
q^2=(p_l-\pnu)^2
\simeq -2E_\nu E_l\left(1-\sin\theta_1\sin\theta\cos\phi
-\cos\theta_1\cos\theta\right).
\label{q2w}
\ee
Regardless of the value of $\phi$, the
right-hand side of Eq.~(\ref{q2w}) is always non-positive.  Therefore $q^2$ in
$\langle{\mathcal M}^2_W\rangle$ of
Eq.~(\ref{invw2}) is either null $(q^2=0)$ or spacelike $(q^2<0)$.
A $W^\pm$ pole is impossible in principle in the charged current case
because it is always true that $q^2\neq M^2_W$.

The invariant amplitude for the neutral current diagram in Fig.~\ref{fd}b
is
\ba
{\mathcal M}_Z&=&-{g_Z^2\over8(q^2-M_Z^2)}\left[\bar{v}_{\bar\nu}\gamma^\mu
(1-\gamma^5)u_\nu\right]\left(g_{\mu\nu}-{q_\mu q_\nu\over M_Z^2}\right)
\left[\bar{u}_f\gamma^\nu(C_V-C_A\gamma^5)v_{\bar f}\right]\nonumber\\
&\simeq&-{g_Z^2\over8(q^2-M_Z^2)}\left[\bar{v}_{\bar\nu}\gamma^\mu
(1-\gamma^5)u_\nu\right]
\left[\bar{u}_f\gamma_\mu(C_V-C_A\gamma^5)v_{\bar f}\right].
\label{invz}
\ea
The flavor of $\nu\bar\nu$ cannot mix.  The type of $f\bar f$ can vary
but the flavor cannot mix.  The values of the neutral vector coupling $C_V$
and axial-vector coupling $C_A$ are given by the GWS (Glashow-Weinberg-Salam)
Model and are tabulated in Table~\ref{cc}.  Applying the usual rules,
the invariant amplitude square can be computed as
\be
\langle{\mathcal M}^2_Z\rangle={1\over4}\left[{g^2_Z\over q^2-M_Z^2}\right]^2
\left\{\left(C_V^2+C_A^2\right)\left[(\pnub\cdot\pf)^2+(\pnu\cdot\pf)^2\right]
+2C_A C_V\left[(\pnub\cdot\pf)^2-(\pnu\cdot\pf)^2\right]\right\}.
\label{invz2}
\ee
The parametrizations
in Eqs.~(\ref{p1})--(\ref{p4}) can be re-applied with the substitutions
$p_1=\pnu$, $p_2=\pnub$, $p_3=p_f$ and $p_4=p_{\bar f}$
The transfer momentum square in the neutral current case is
\be
q^2=(\pnu+\pnub)^2\simeq 2(\pnu\cdot\pnub)=2E_\nu E_{\bar\nu}
[1-\cos(\theta_1+\theta_2)].
\label{q2z}
\ee
According to Fig.~\ref{fd}b, $0\le\theta_i\le\pi/2$ for $i\in\{1,\,2\}$ so
that $q^2\ge0$ in Eq.~(\ref{q2z}).
Therefore a $Z^0$ pole is possible for $\langle{\mathcal M}^2_Z\rangle$
in Eq.~(\ref{invz2}) at $q^2=M^2_Z$.
On the average, $\langle\cos\phi\rangle=0$ and
$\langle\cos^2\phi\rangle=1/2$.
The remaining kinematic factors in $\langle{\mathcal M}^2_Z\rangle$
of Eq.~(\ref{invz2}) can be calculated as
\ba
(\pnu\cdot\pf)^2&=&E^2_\nu\,E^2_f\left(1-2\cos\theta_1\cos\theta
+{1\over2}\sin^2\theta_1\sin^2\theta+\cos^2\theta_1\cos^2\theta\right),
\label{kz1}\\
(\pnub\cdot\pf)^2&=&E^2_{\bar\nu}\,E^2_f\left(1-2\cos\theta_2\cos\theta
+{1\over2}\sin^2\theta_2\sin^2\theta+\cos^2\theta_2\cos^2\theta\right).
\label{kz2}
\ea
Letting $\enu=\enub$ leads to $\theta_1=\theta_2$ which in turn gives
$(\pnu\cdot\pf)^2=(\pnub\cdot\pf)^2$
so that the form of Eq.~(\ref{invz2}) is greatly simplified.  Letting
$\enu=\enub$ has an additional advantage that
the intensities of $\nu_e$, $\nu_\mu$ and $\nu_\tau$ in the neutrino beam
will match those of $\bar\nu_e$, $\bar\nu_\mu$ and $\bar\nu_\tau$
in the antineutrino beam despite of neutrino oscillation as long as
the original neutrino-antineutrino beams have the same intensities and
the matter effect of the earth is either negligible or canceled by
symmetry.

The cross section formula of the reaction $1+2\to3+4+\cdots+n$ can be
calculated by the Fermi's Golden Rule~\cite{griffiths},
\ba
d\sigma&=&\langle{\mathcal M}^2\rangle
{S\over4\sqrt{(p_1\cdot p_2)^2-(m_1m_2)^2}}
\left[
\left({d^3{{\mathbf p}_3}\over(2\pi)^3 2E_3}\right)
\left({d^3{{\mathbf p}_4}\over(2\pi)^3 2E_4}\right)\cdots
\left({d^3{{\mathbf p}_n}\over(2\pi)^3 2E_n}\right)
\right]\nonumber\\
&&\quad\times(2\pi)^4\delta^4(p_1+p_2-p_3-p_4-\cdots-p_n),
\label{ds}
\ea
where $S$ is a statistical factor that includes a factor of $1/j\!$ for each
type of $j$ identical outgoing particles.  For the Feynman diagrams in
Fig.~\ref{fd}, $S=1$ and $n=4$.
The differential cross section for the neutral current diagram can be
obtained from Eq.~(\ref{ds}) after performing integrations over ${\mathbf p}_3$
and ${\mathbf p}_4$, 
\ba
{d\sigma\over d\Omega}&=&{1\over(8\pi)^2}\,
{\langle {\mathcal M}_Z^2\rangle \over\enu\enub[1-\cos(\theta_1+\theta_2)]}
\nonumber\\
&&\qquad\times{E_f\over\enu+\enub-\cos\theta
\sqrt{(\enu\sin\theta_1-\enub\sin\theta_2)^2
+(\enu\cos\theta_1+\enub\cos\theta_2)^2}}.
\label{dss}
\ea
Eq.~(\ref{dss}) is useful for checking pathology in the theory.  In this case,
there is no pathology related to the cross section formula.  It is
checked that $d\sigma/d\Omega\propto\langle{\mathcal M}^2_Z\rangle$.
Although the exact form of the cross section formula is unimportant for the
the purpose of this work, Eq.~(\ref{dss}) is useful for calculating the
relative
distributions of the outgoing particles once the infinity of $1/[q^2-M^2_Z]$
at the $Z^0$ pole is canceled out by division.  At the $Z^0$ pole, the cross
section in Eq.~(\ref{dss}) is infinite and the interaction probability is 1.
This way, the flux of annihilation-induced radiations is directly
proportional to the intensities of the $\nu\bar\nu$ beams.  For the purpose
of designing an efficient neutrino weapon,
Eq.~(\ref{q2z}) used in $q^2=M^2_Z$ is the key result of this section.
The branching ratios of various annihilation-induced products
$f\bar f$ can be calculated by comparing $\langle{\mathcal M}^2_Z\rangle$
using Eq.~(\ref{invz2}). In the
case $\enu=\enub$, the branching ratios are particularly simple and
can be obtained from ratios of $C_V^2+C_A^2$ alone.

Finally the earlier claim that the width term $iM\Gamma$ can be eliminated
from the final result by special relativity needs to be explained.
Historically the width is incorporated in the scattering amplitude $f(E)$ in
non-relativistic quantum mechanics as in~\cite{peskin,landau}
\be
f(E)\propto{1\over E-E_0+i\Gamma/2}.
\label{fE}
\ee
It is observed in low energy inelastic scattering experiments that an
incoming particle form a compound nucleus with the nucleons of the target
nucleus as an intermediate state.  The discrete energy levels $E_0$ of this
compound nucleus give rise to resonances.  If the compound nucleus is unstable,
the resonance is modified as $E_0\to E_0-i\Gamma/2$.  It is the so-called
Breit-Wigner resonance.  In experiments, $E_0$ and $\Gamma$ are extracted from
partial wave analysis.  In hadron physics, $\Gamma$ is modeled as a function
of energy $\Gamma\to\Gamma(W)$, where $W$ is the energy in the
center-of-momentum (CM) frame, along with other model dependent
parameters~\cite{manley}. With minor adjustments, the Breit-Wigner formula
is phenomenologically robust.  Sometimes there are disagreements among
different analyses on some of the parameters.  An
example is the $E_{0+}$ parameter in the cross section of the $\eta$
photoproduction of $S_{11}(1535)$~\cite{balandina,dugger,renard,krusche}.
In this case, the discrepancy is thought to have come from different
experimental biases and the difficulties in resolving nearby resonances.
So far the validity of Eq.~(\ref{fE}) is not questioned.
Although the form of Eq.~(\ref{fE}) has been discussed in many quantum
mechanics textbooks~\cite{landau}, the $iM\Gamma$ term in the propagator of
Eq.~(\ref{prop2}) in quantum field theory is not derived from first
principle but is borrowed from quantum mechanics.
Reference~\cite{peskin} contains a heuristic explanation of Eq.~(\ref{prop2})
as
\be
{1\over p^2-M^2+iM\Gamma}\approx
{1\over 2E_{\bf p}(p^0-E_{\bf p}+i(M/E_{\bf p})\Gamma/2)}.
\label{heur}
\ee
In Eq.~(\ref{heur}), $p^0$ is the pole and $E_{\bf p}$ is the
energy of the particle.  The factor $M/E_{\bf p}$ is the reciprocal of the
Lorentz factor $1/\gamma$ that accounts for the relativistic
correction of the
lifetime of the particle $\tau=\gamma\tau_0$.  The right-hand side of
Eq.~(\ref{heur}) is a combination of Eq.~(\ref{fE}) with the replacement
$\Gamma\to\Gamma/\gamma$ and a factor of $1/2E_{\bf p}$ that commonly appears
in relativistic quantum mechanics and quantum field theory.  Since the
relativistic effect is already included by virtue of incorporating
$1/\gamma$ on the right-hand side of Eq.~(\ref{heur}),
$\Gamma$ on the left-hand side can be taken as a constant much like the rest
mass $M$ is a constant.  The implicit Lorentz factor $\gamma$ on
the right hand side of Eq.~(\ref{heur}) is frame dependent.  The
left hand side is manifestly frame independent.  This ambiguity is a proof that
Eq.~(\ref{heur}) is not a theory but a phenomenological patch.
Since Eq.~(\ref{heur}) is borrowed from Eq.~(\ref{fE}), it is reasonable to
assume that the validity of the former is based on the satisfaction of
the same conditions of the latter.  The basic assumptions of
Eq.~(\ref{fE}) are the target nucleus and compound nucleus at rest in the
lab frame and a total width $\Gamma$ measured in the same rest frame.
In high energy scattering experiments such as $e^+e^-\to Z^0$ and
$p\bar p\to Z^0$, the equivalent of a target nucleus can be taken to be
the center-of-mass of the in-state.  So far
$e^+e^-$ and $p\bar p$ collider experiments are always arranged in such
a way that the beams are colliding head-on with equal and opposite
momenta.  The CM frame is the natural reference
frame for the data analysis of such collider experiments.  The 3-momentum of
$Z^0$ in the CM frame is zero (${\bf q}=0$) so that $Z^0$ is at rest in this
frame.  Additionally the center-of-mass frames of the in and out states
are also at rest because ${\bf q}=0$.  Since all the relevant systems are at
rest in the CM and lab frames, it is reasonable to expect that the widths at
the $Z^0$ pole measured by collider experiments agree with the non-relativistic
Breit-Wigner formula.  In neutrino weapon
applications, it is seldom practical to shoot the $\nu\bar\nu$ beams head-on.
In the lab frame where the $\nu\bar\nu$ beams are colliding at oblique angles,
the momenta of center-of-mass of $\nu\bar\nu$, $Z^0$ and $f\bar f$
are non-zero,
${\mathbf q}={\mathbf p}_1+{\mathbf p}_2={\mathbf p}_3+{\mathbf p}_4\ne0$.
In order to understand the phenomenology of the $Z^0$ pole in the ${\bf q}\ne0$
case, the focus is directed to the type of collisions in that a resonance is
moving relativistically with respect to the target.  According to
Eq.~(\ref{q2z}), the $Z^0$ pole is possible
in a stationary target when the beam energy reaches $4\times10^6$~GeV for
a positron beam on an electron target or $2\times10^3$~GeV for a anti-proton
beam on a proton target.  Such experiments do not exist.
In the lower energy regime, nuclear physics experiments often shoot electron
or photon beams at stationary targets.  The resonances are usually excitations
of proton so that the target and the resonance are at rest
with respect to each other.  In rare high energy reactions
$e^+e^-\to Z^0Z^0$~\cite{barato,abdallah,achard,abbiendi}
and $p\bar p\to Z^0Z^0$~\cite{abazov}, the condition ${\bf q}\ne0$ is obtained
automatically by virtue of integrating over ${\bf p}_3$ and ${\bf p}_4$ in
the calculation of the cross section formula.  However $Z^0$ is the out-state
in this case and not the propagator so that the
Breit-Wigner formula does not apply.  To the best of the author's
knowledge, there is no scattering experiment that satisfies the condition
of ${\bf q}\ne0$ at the $Z^0$ pole at the time of the writing of this paper.

At the $Z^0$ pole, most of the energy of the in-state is converted to the
$Z^0$ mass so that the residual kinetic energy is relatively small.
The $Z^0$ propagator is approximately at rest in the lab frame even though its
momentum is non-zero ${\bf q}\ne0$.  Electron and neutrino, on the other hand,
have very small mass so that a relatively small amount of transfer
momentum ${\bf q}$ will cause them to move relativistically.  $Z^0$ is a
virtual particle.  Its influence is limited to a short time between the
in and out states.  As far as the in and out states in their center-of-mass
rest frames are concerned, a moving $Z^0$ of total width $\Gamma$ is
effectively the same as a hypothetical stationary $Z^0$ of width
$\Gamma/\gamma$.  The advantage of modeling the propagator as a
stationary $Z^0$ in the center-of-mass rest frame of the in-state is
that this condition satisfies the assumption of the non-relativistic
Breit-Wigner formula in that both the target and resonance are at rest with
respect to each other.  The similar argument can be repeated for the out-state.
Following the same reasoning of Eq.~(\ref{heur}), the propagator in the case of
${\mathbf q}\ne0$ is taken to be
\be
{1\over(q^2-M^2_Z+2im\Gamma)},
\label{prop3}
\ee
where $m\in\{m_\nu,\,m_f\}$.  The factor of 2 in the width term $im\Gamma$ of
Eq.~(\ref{prop3}) comes from the fact that there are 2 particles of mass $m$
in the in and out state.  The subtlety of Eq.~(\ref{heur}) is that the
argument is crafted first in a preferred reference frame (such as the lab
frame) and then generalized to be frame independent.  Eq~(\ref{prop3})
exploits the same strategy by incorporating a Lorentz factor $\gamma=q_0/(2m)$
in the center-of-mass rest frame of the in-state (out-state).  $q_0$ as in
$q=(q_0,\,{\bf q})$ is canceled from Eq.~(\ref{prop3}) by the
factor $1/(2q_0)$.  With Eq.~(\ref{prop3}),
$\langle{\mathcal M}^2_Z\rangle$ for ${\mathbf q}\ne0$ can be re-evaluated
by averaging the invariant amplitude squares computed with the new total
widths corresponding to the in and out states,
\ba
\langle{\mathcal M}^2_Z\rangle&\propto&
{1\over2}\lim_{m_\nu\to0}
\left\{{1\over\left(q^2-M^2_Z\right)^2+4m^2_\nu\Gamma^2}+
{1\over\left(q^2-M^2_Z\right)^2+4m^2_f\Gamma^2}\right\}\nonumber\\
&\simeq&{1\over2}\left\{{1\over\left(q^2-M^2_Z\right)^2}\right\}.
\label{invz3}
\ea
The last step of Eq.~(\ref{invz3}) is motivated by
$1/m^2_\nu\ll 1/m^2_f$.  A third term constructed from the rest frame of
$Z^0$ is not needed because it is redundant.
In actuality, $m_\nu$ is small but non-zero.  Assuming that
$m_\nu\sim0.1$~eV, $M^4_Z/m^2_\nu\Gamma^2\sim10^{27}$.  Instead of an infinite
cross section as claimed earlier, the reaction rate at the $Z^0$ pole is
only 27 orders of magnitude larger than that of the inverse beta
decay.  The improvement is adequate to make
the idea of a neutrino counter nuclear weapon practical with reasonably
intense $\nu\bar\nu$ beams.  Therefore the invariant amplitude square in
Eq.~(\ref{invz2}) is a good approximation up to a factor of 2.

\section{Simulation Results}
The exact details of the designs of thermonuclear weapons are classified.
However the basic principles of the operation of various types of nuclear
weapon are unclassified and freely available through
public information~\cite{bernstein}.
Generally speaking, a modern
nuclear weapon is made up of a fission bomb (the primary stage) and a fusion
bomb (the secondary stage).  The primary stage typically has a plutonium
core surrounded by a tamper, a neutron reflector and a set of chemical
explosive lenses on the outermost layer. The tamper slows down the explosion
just enough to increase the efficiency of the chain reactions.  The
neutron reflector increases the criticality of the fissile so that the
efficiency of the explosion is further increased.  In some designs (such as
the W87 warhead), the tamper and
neutron reflector seem to be combined in one layer made of beryllium.
The secondary stage is a spherical or
cylindrical complex of solid lithium deuteride (fusion fuel) interlaced with
layers of uranium walls and the so-called
``spark plug'' in the center.  A spark plug is a fissile lining that enhances
compression.  The sequence of explosions begins with the simultaneous
ignition of the chemical explosive lenses in the primary stage to compress
the sub-critical plutonium core to reach super-criticality by implosion.  The
heat and pressure generated by the fission bomb in the primary stage
compress the fusion fuel in the secondary stage.  Neutrons from the
fission bomb convert the lithium-6 isotopes in solid lithium deuteride
into tritium
($^6_3{\rm Li} + {\rm n}\to ^4_2{\rm He} + ^3_1{\rm H} + 4.8$~MeV).
Tritium and deuterium fuse to create a hydrogen bomb explosion.  In a
thermonuclear
weapon, most of the energy is generated by fusion in the secondary stage.
A sketch of the W87 warhead found online~\cite{w87} is the source of the
geometry used in the simulations of this work.  More detailed discussions of
thermonuclear weapon designs can be found on the same
archive~\cite{design}.

Neutrinos cannot be shielded and can reach a nuclear weapon anywhere on earth
or in space as long as the location is specified and if the neutrino weapon
can focus the high energy $\nu\bar\nu$ beams on the target accurately.
High energy radiations produced by $\nu\bar\nu$ annihilation
cause neutron spallation inside the fissile material that in turn initiate
fission reactions.  Fissions create heat.  When heat is generated in sufficient
quantity, it ignites the chemical explosive lenses surrounding the plutonium
core in the primary stage of the thermonuclear warhead.  This way
$\nu\bar\nu$ annihilation provides a means to detonate a nuclear weapon
remotely.  The critical question is whether these
annihilation-induced radiations can be generated efficiently and cheaply
to make the idea practical.
To this end, the simulation tool \texttt{MCNPX2.5.0}
is used to calculate the heat deposition by fissions from high energy
external sources at $E_f=45$~GeV, approximately half the value of $M_Z$.
\texttt{MCNPX2.5.0} is an integration between \texttt{MCNP4B} and
\texttt{LAHET2.8}.  It uses a very old version of \texttt{FLUKA} 
(\texttt{FLUKA87}) as a hadron generator above the INC (intranuclear
cascade) region ($E>10$~GeV).
The manual of \texttt{MCNPX2.5.0} explicitly states that it does not include
any nuclear reactions for muon.  Muon-induced neutron spallation has
been observed experimentally and is already included in the more recent
versions of \texttt{FLUKA}.  A newer version of \texttt{FLUKA} is planned to be
integrated in the next release of \texttt{MCNPX} so that this problem will
hopefully be fixed in due time.  Another problem of \texttt{MCNPX2.5.0}
inherited from \texttt{LAHET} is that
not all particle types can be used as primaries.  So far it is found that
\texttt{MCNPX} will crash when $\tau$ is used as a primary.  During the
course of this work, it is also discovered that \texttt{MCNPX} has a strange
bug when $\pi^0$ is used as a primary.  The bug is already reported to the
\texttt{MCNPX} team.  In principle, spallation neutrons induced by $\mu^\pm$,
$\tau$ and $\pi^0$ can be simulated by the latest version of \texttt{FLUKA}
and then input into \texttt{MCNPX} to calculate the fission deposition
energy.  Unfortunately the license agreement of \texttt{FLUKA} restricts
the software to non-weapon-related use.  Since the purpose of the present
work is merely a feasibility study and not a detailed weapon design, the
absence of $\mu^\pm$, $\tau$ and $\pi^0$ as primaries in the
simulations will not affect the conclusion.  The remaining
particle types used as primaries are $e^\pm$, $\pi^\pm$, $K^\pm$,
$K_L$ and $K_S$.  Of course neutrinos can also be annihilation products.
But they will not be a useful source of weapon related radiations and
will not be simulated. Despite of the shortcomings aforementioned,
the advantages of \texttt{MCNPX} are a complete set of nuclear
libraries, very good neutron simulations and a card to calculate
fission energy deposition.  Therefore \texttt{MCNPX}
is still the best simulation tool for the purpose of the present study.
In the absence of detailed designs of nuclear weapons and a
complete simulation package, the present feasibility study is meant to be
an order of magnitude estimation of fission energy deposition
from annihilation-induced products.

\begin{figure}
\includegraphics[scale=0.6]{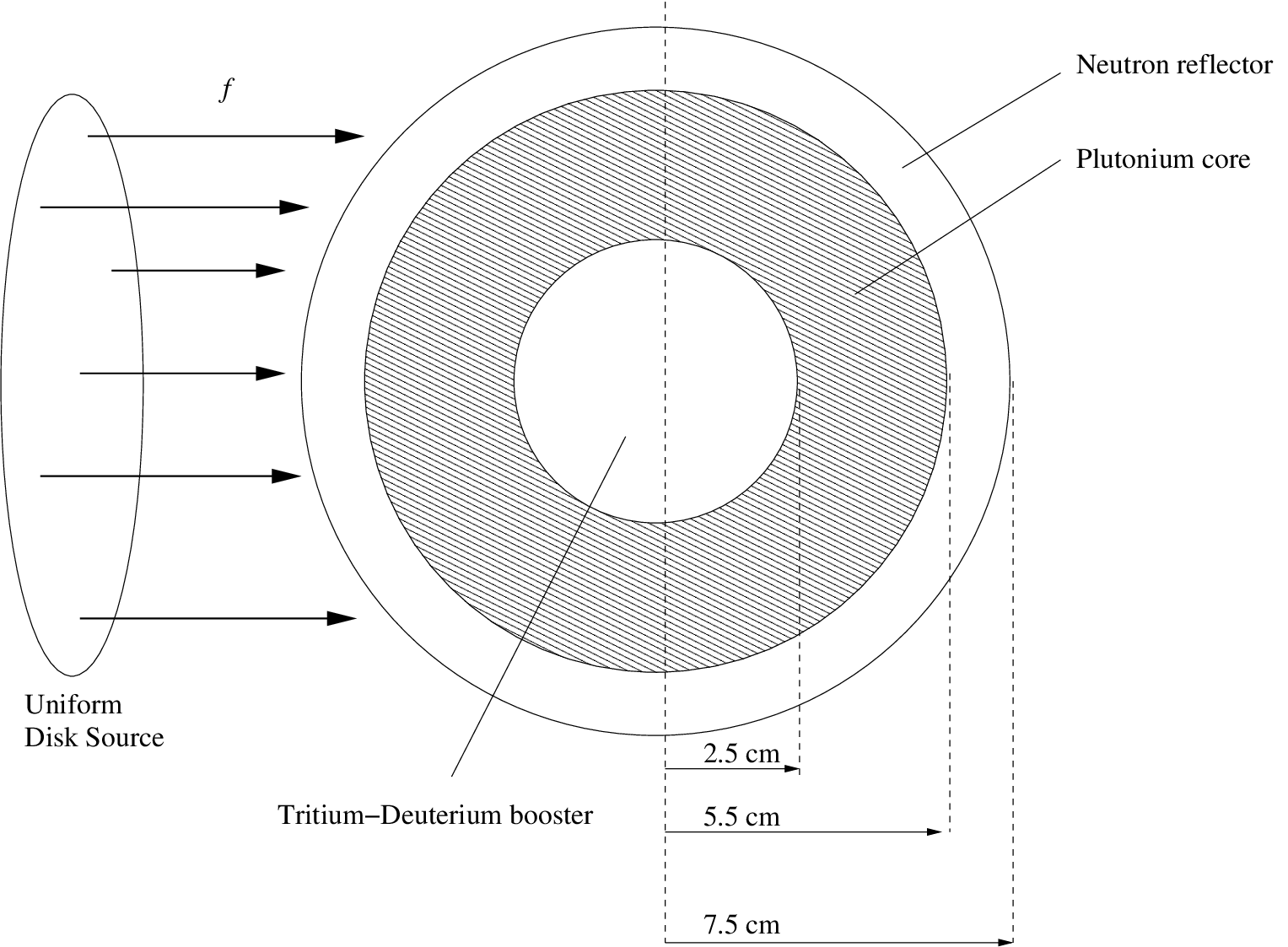}
\caption{\label{core}
A mock-up of the primary stage of the W87 thermonuclear warhead with
the explosive lenses removed.  The core is made out of plutonium alloy
(99.4\%  $^{239}{\rm Pu}$ and 0.6\% natural gallium).  The density of
plutonium alloy is 15.86~g/cc.  The neutron reflector
is made of beryllium.  The density of Be is 1.85~g/cc.
The tritium-deuterium booster has a density
of 0.00175~g/cc.  The source is
$f\in\{e^\pm,\,\pi^\pm,\,K^\pm,\,K_L,\,K_S\}$.}
\end{figure}

The geometry of the primary stage of the W87 thermonuclear warhead used in the
\texttt{MCNPX} simulations is illustrated in Fig.~\ref{core}.  Due to the
spherical symmetry of the geometry, a uniform disk source is adequate
to simulate random radiations on the core.  On a 2~GHz \texttt{Core Duo 2}
processor, a particle history of an external $e^\pm$ takes 2-3 hours to
complete on one core.  Primaries of other particle types may take only 2-3
minutes to finish.  In order to be fair in the comparisons of all particle
types, the same number of
particle history (\texttt{NPS}) is used for all of the simulations, namely
$\texttt{NPS}=100$.  Table~\ref{ec} summarizes the annihilation-induced
fission energy deposition on the plutonium core of a mocked-up W87 warhead.
The unit of fission energy deposition output by \texttt{MCNPX} is
MeV/g/$p$ (where $p$ stands for ``primary'') and is converted to MeV/$p$
by multiplying
the output with the mass of the plutonium core $1.0015\times10^4$~g.
The energy deposition on the human body is simulated by shooting the
primaries from a point source inside a rectangular parallelepiped of
$200\times100\times100\,{\rm cm}^3$ made with a material of similar chemical
composition of
the human body as shown in Table~\ref{human}.  The simulation results
of annihilation-induced energy deposition on the human body is summarized in
Table~\ref{eh}.  The unit of energy deposition in \texttt{MCNPX} is
MeV/g/$p$ and is converted to Gray per primary (Gy/$p$) where Gray is
defined as J/Kg.  The large errors in Tables~\ref{ec} and \ref{eh} are
partly due to varying slant depths when the primaries intersect
different parts of the targets.

A typical accelerator such as MINOS can produce neutrinos (93\% $\nu_\mu$,
6\% $\bar\nu_\mu$ and 1\% $\nu_e+\bar\nu_e$) from a beamline of
$2\times10^{13}$~protons/s.
High energy protons interact in a target on the NUMI beamline to produce
pions and kaons, which decay into muons and muon neutrinos
($\pi^+\to\mu^+\nu_\mu$).  Muons further decay into electrons and neutrons
($\mu^+\to e^+\nu_e\,\bar\nu_\mu$).
The energy spectrum peaks at approximately 3~GeV with a long high-energy tail
extending to 120~GeV~\cite{minos}.  Assuming a hypothetical situation
in that the peak of the neutrino spectrum is tuned to $M_Z/2$ for the
$Z^0$ pole, much of energy is still lost to smearing and the production
of by-products.  An accelerator of the size of
Fermilab is not practical for neutrino weapon applications because of
the problem of maneuverability and the prohibitive cost.  According to
Table~\ref{ec}, the typical fission energy deposition is in the
$10^4$~MeV/$p$ range.  Assuming a very optimistic estimate of 5\% efficiency
for energy transfer, neutrinos produced by an optimized version of MINOS and
its antineutrino counterpart will create enough annihilation-induced
radiations to convert the plutonium core in Figure~\ref{core} into a
1--10~kW reactor.  Suppose that the chemical
explosive around the plutonium core is C4 which is 91\% RDX
(cyclotrimethylene-trinitramine) and has a flash point at $234^\circ$.
Reference~\cite{sugawara} estimates the heat capacity of $^{239}$Pu as
$6.557\times10^{11}$~MeV/g$\cdot$K while Reference~\cite{groper} shows a
more detailed temperature dependent heat capacity profile for both pure and
alloyed plutonium as 30--40~J/mol$\cdot$K at 20--250~$^\circ$C.
As an approximation, this work takes the heat capacities of the plutonium alloy
and beryllium to be $8\times10^{11}$~MeV/g$\cdot$K and
$1.14\times10^{13}$~MeV/g$\cdot$K respectively.  A MINOS-induced
1--10~kW heater can detonate a plutonium core in Figure~\ref{core}
in 100--1000~seconds.  Long heating time will likely
lead to the heating of the area surrounding the
plutonium core.  Before the temperature in the primary stage
reaches the flash point of the
explosive, the polyethylene filler may melt and the supporting
structure may expand. If the position of the fissile shifts relative to its
surrounding, the effectiveness of the explosive lenses may be affected in
such a way that a fizzile (not a full scale nuclear explosion) occurs.
In case the nuclear weapon is moving, it is tactically advantageous to minimize
the heating time because of the technical difficulty of tracking.  For these
reasons, the heating time should be limited to approximately 1~s to detonation.
The main disadvantage of a conventional accelerator is not just
the long heating time but also its size and cost.  Fortunately
recent breakthroughs on tabletop accelerators may one-day be the solution
for all these problems.  Plasma accelerators utilizing the laser-driven
Wakefield can achieve an electric field gradient up to 270~GeV/m depending on
the electron density in the plasma.  The electron beam  produced by
Laser Wakefield Acceleration (LWFA) is stable and collimated~\cite{faure}.
LWFA has some potential limitations.  A study
shows that electric field breakdown occurs at $13.8\pm0.7$~GeV/m in some
dielectric~\cite{thompson}.  In one report,
a mono-energetic electron beam up to 1~GeV driven by a 40~TW laser in a
3.3~cm-long hydrogen-filled capillary discharge wave guide has
been successfully demonstrated~\cite{gonsalves}.  The peak current achieved
is up to 300~A ($2\times10^{21}$~electrons/s).  At the
time of the writing of this paper, the highest peak current published by
another report is 100~pC/10~fs or 10~kA~\cite{schroeder,tilborg}.  Results
of a positron laser-plasma accelerator is recently published~\cite{wang}
so that all of the essential components for the construction of $\nu\bar\nu$
beams are already in place.  Assuming a typical femtosecond laser
pulse rate of 30~kHz at the current level of technology, LWFA $\nu\bar\nu$
beams of 1--10\% efficiency for energy transfer
can detonate the plutonium core in Figure~\ref{core} in 50--500~s.

The radiations produced by the neutrino counter nuclear weapon can also
irradiate human
beings.  The lethal dose for a typical human adult is 10-20~Grays.
According to Table~\ref{eh}, the dosage from annihilation-induced
radiations is in the $10^{-14}-10^{-12}$~Gy/$p$ range.  Assuming
an average dosage of $10^{-13}$~Gy/$p$ per beam and the same parameters
of the LWFA $\nu\bar\nu$ beams above, a neutrino tactical weapon takes
10--100~s to achieve the lethal dose per person.

\section{Engineering}
Neutrino beams are typically produced by colliding particles on suitable
targets in a high energy accelerator such as those in MINOS and CNGS to
create pions and kaons which decay into neutrinos, small amount of
anti-neutrino and hadrons.  As it is shown above, a Fermilab size
accelerator is not practical for
the purpose of a neutrino counter nuclear weapon because a counter weapon has
to be small, mobile and must have aiming capability.  Laser wakefield
accelerator (LWA) is small and agile.  It has already succeeded in producing
1~GeV electrons in a space of less than 1~cm long so that it has potential to
be the accelerator of choice for a neutrino counter nuclear weapon.  In LWA,
an intense laser femtosecond laser pulse creates a bubble of ion cavity that
follows the laser pulse.  Electrons create a wakefield in the ion cavity which
in turn accelerate other electrons.  MCNP simulation shows that each neutrino
anti-neutrino annihilation at the $Z_0$ pole deposits about 104~MeV of energy
on the plutonium core.  Taking into consideration the heat capacity of
plutonium and the temperature of the flash point of the explosive surrounding
the plutonium core, a LWA based neutrino counter nuclear weapon operated on a
1~TW laser can in principle detonate a nuclear weapon in a few
minutes.  The next generation of femtosecond laser is projected to reach
Peta Watts.  It is possible to create a LWA based neutrino counter nuclear
weapon that can detonate a nuclear warhead instantaneously.

Currnetly LWA makes an electron beam by accelerating electrons through the ion
bubble in a plasma.  A positron beam is made by an entirely different technique
which uses a powerful laser to bombard a target (e.g. gold foil) to generate
and accelerate positrons at the same time.  When electron and positron beams
hit suitable targets to create pions and kaons, they
decay into neutrinos and anti-neutrinos in similar ways as those in MINOS and
CNGS.  This conventional method of generating neutrino and anti-neutrino beams
is not very energy effiicient.  In MINOS and CNGS, a 120~GeV proton beam
produces 12-14~GeV neutrinos.  The mass of $Z_0$  is about 91G~eV.  Depending
on the angle $\theta$ between the neutrino and anti-neutrino beams the
transfer momentum square is
$q^2\simeq 2E_\nu E_{\bar\nu}(1-\cos\theta)$ according to Eq.~(\ref{q2z}).
It can be seen that the neutrino and anti-neutrino beam energies need to be
comparable to the $Z_0$ mass. Using conventional method to generate neutrino and
anti-neutrino beams, the LWA electron and positorn beam energies have to be
in the order of 1~TeV.  Assuming an accelerating gradient of 1 GeV/cm for LWA,
the length of the accelerator must be more than 100~m which is beginning to
challenge the assumption that a LWA based neutrino counter nuclear weapon is
compact.  Additonally pion and kaon decays do not produce pure neutrino and
anti-neutrino beams.  Mixture of neutrino species in a beam adds to the
complexity of engineering design.  Therefore a non-conventional technique for
generating neutrino and anti-neutrino beams is needed.

CERN has been experimenting with the idea of beta beam which involves
accelerating heavy ions that are beta emittors~\cite{wildner}.
Low Q value neutrino and anti-neutrino emitters are $^6{\rm He}$ and
$^{18}{\rm Ne}$.  High Q value emitters such $^8{\rm Li}$ and $^8{\rm B}$
can provide neutrinos and anti-neutrinos at 3.5 times the energies compared to
low Q value emitters.  For a Lorentz factor of $\gamma=100$, the neutrino and
anti-neutrino beta beams generated so far can reach the energy range of
atmospheric neutrinos (2~TeV to 200~TeV) which is more than adequate to
create a $Z_0$ pole.  Currently beta beam is made by conventional RF
accelerators.  Reseach has been conducted on using LWA to accelerate heavy
ions for more than a decade~\cite{hiroaki}.  Currently Trident Laser Facility
at LANL has achieved 100~MeV to 1~GeV proton and heavy ion energy through
laser-driven acceleration by hitting a suitable target with a table top
laser~\cite{jung}.  However beta emitters may not be accelerated using the
laser bombardment technique.  Instead it is hoped that technology may soon
allow the integration of LWA and beta beam for neutrino counter nuclear
weapon applications.

Some of the key challenges of a LWA based neutrino counter nuclear weapon are
plasma instability and neutrino beam quality.  Plasma instability in LWA is
being actively researched~\cite{huntington}.  Physicists are now using LWA to
build x-ray lasers so that confidence on the LWA beam quality is
reasonable~\cite{corde}.  In the conventional colliders such as Tevatron and
LHC, the proton and anti-proton  beams are squeezed just before the collision
to increase the particle density and hence collision probability.  In a
neutrino counter nuclear weapon, squeezing neutrino and anti-neutrino beams
prior to the neutrino anti-neutrino annihilation is not a possibility.
Therefore the neutrino and anti-neutrino beams have to be produced very
collimated and dense from the start to increase of probability of collisions.
The beam energies also need to be very fine tuned to improve the efficiency of
neutrino anti-neutrino annihilation at the $Z_0$ pole.  Therefere many
technological problems still need to be solved before a LWA beta beam based
neutrino counter nuclear weapon is feasible.

\section{Discussions and Conclusion}
An order of magnitude calculation based on a mocked-up design of the
W87 warhead and a set of preliminary \texttt{MCNPX} simulations shows that
radiations from $\nu\bar\nu\to f\bar f$ reactions at the $Z^0$ pole is capable
of detonating a nuclear weapon remotely.  The theory of the neutrino weapon
capitalizes on the large cross section of the $Z^0$ pole at oblique
angles.  If the assumptions of relativistic arguments leading up to
Eq.~(\ref{invz3}) are incorrect and that the width is not substantially reduced
as claimed, the efficiency of a neutrino weapon will seriously suffer to
the point of being impractical.  Assuming that the theory is correct,
the efficiency of the neutrino weapon will largely depend on our
ability to control the parameters in Eq.~(\ref{q2z}) to satisfy the condition
$q^2=M^2_Z$ exactly.
Since $\langle{\mathcal M}^2_Z\rangle\sim1/[q^2-M^2_Z]^2$ in
Eq.~(\ref{invz2}), an agreement between $\enu$ and $\enub$ with $M_Z$ to 4
significant figures implies a $10^{12}$ increase in the cross section and so
on.  In reality,
there will always be a smear of neutrino energy around the $Z^0$ pole.
Too much smearing will decrease the efficiency of the neutrino weapon.
The reaction $\nu\bar\nu\to\nu\bar\nu$ also decreases the efficiency of the
neutrino weapon but this effect is predictable and inconsequential.
The heat transfer from the plutonium core to the chemical explosives can
in principle be reduced by thermal insulation or by changing the design
to a gun-type trigger mechanism.  These fixes will most likely reduce the
efficiency of a nuclear weapon and may not protect the weapon from being
detonated remotely.  The neutrino counter nuclear weapon
can also be used as a tactical weapon.  The advantage of a neutrino
tactical assault
weapon is that it can hit hard-to-reach places such as deep underground
concrete bunkers and caves in mountainous areas.
Since neutrinos can travel across the globe in the speed of light, a
neutrino weapon does not allow the time for an early warning system.  Since
neutrinos cannot be shielded, a neutrino weapon is in principle
non-defensible.  A preliminary analysis of the capability of a neutrino weapon
given above is based on the experimental designs of plasma
accelerators. The order of magnitude estimates show that the current level of
technology is already meeting the minimum hardware requirements.
It is foreseeable that high quality and high intensity $e^+e^-$
beams will be made available by LWFA and similar technologies soon.
The remaining technological challenge is to engineer collimated
mono-energetic high intensity $\nu\bar\nu$ beams using beta emitters.
Accelerator R\&D will be the most critical part
of the development of a neutrino weapon.  Experimental verification of the
near-singularity of the $Z^0$ pole at oblique angles is also very
important.  Since neutrino propagation through the Earth is affected by
neutrino oscillation and matter effect, basic neutrino physics research is
needed to measure the physical parameters accurately.  In order to obtain
maximal collision efficiency, the $\nu\bar\nu$ beams must be focused on a small
spatial region.  Due to the relatively low rate of lethal radiation on a
human body, a neutrino tactical weapon is a unlikely candidate for a weapon
of mass destruction.  At the end, it is hoped that
the neutrino counter nuclear weapon will make nuclear weapons obsolete
so that the goal of non-proliferation is achieved peacefully.

\begin{acknowledgments}
The author thanks the Chinese University of Hong Kong for hosting his visit
during the period of this research.  V. Taranenko, G. W. McKinney, M. Sher,
W. M. Morse, W. Simmons and C. G. R. Geddes have graciously provided helpful
criticisms and discussions that improve the quality of the paper at various
stages.
\end{acknowledgments}

\begin{table}
\caption{Neutral vector and axial vector couplings in the GWS
Model~\cite{griffiths}.}
\begin{ruledtabular}
\begin{tabular}{rrr}
f & $C_V$ & $C_A$ \\
\tableline
$\nu_e$, $\nu_\mu$, $\nu_\tau$ & $1\over2$
& $1\over2$ \\
$e^-$, $\mu^-$, $\tau^-$       & $-{1\over2}+2\sin\theta_w$
& $-{1\over2}$ \\
$u$, $c$, $t$                  & ${1\over2}-{4\over3}\sin\theta_w$
& $1\over2$ \\
$d$, $s$, $b$                  & $-{1\over2}+{2\over3}\sin\theta_w$
& $-{1\over2}$
\end{tabular}
\end{ruledtabular}
\label{cc}
\end{table}

\begin{table}
\caption{Fission energy deposition $E_c$ from $\nu\bar\nu$ annihilation-induced
products used as primaries $p$ on the plutonium core of the primary stage of
a mocked-up W87 thermonuclear warhead.  The geometry is illustrated
in Fig.~\ref{core}.  The number of particle history is 100.  The energy of
the primaries is $E_f=45$~GeV.}
\begin{ruledtabular}
\begin{tabular}{lr}
$p$ & $E_c$ \\
    & $(10^{4}\,{\rm MeV}/p)$ \\
\tableline
$e^\pm$    &  $2.30\pm0.36$ \\
$\pi^\pm$  &  $5.80\pm0.59$ \\
$K^\pm$    &  $2.59\pm0.39$ \\
$K_L$      &  $2.11\pm0.34$ \\
$K_S$      &  $1.58\pm0.25$
\end{tabular}
\end{ruledtabular}
\label{ec}
\end{table}

\begin{table}
\caption{Approximate chemical composition of the human body.  The average
mass and density of a human body is taken to be 75~Kg and 1.4~g/cc
respectively.}
\begin{ruledtabular}
\begin{tabular}{lr}
Element & Relative Atomic Composition \\
\tableline
O   &  65 \\
C   &  18 \\
H   &  10 \\
N   &  3  \\
Ca  &  1.5 \\
P   &  1.2 \\
K   &  0.2 \\
S   &  0.2 \\
Cl  &  0.2 \\
Na  &  0.1 \\
Mg  &  0.05
\end{tabular}
\end{ruledtabular}
\label{human}
\end{table}

\begin{table}
\caption{Energy deposition $E_h$ from $\nu\bar\nu$ annihilation-induced
products used as primaries $p$ on the human body.  The chemical composition of
the human body is given in Table~\ref{human}.  The number of particle history
is 100.  The energy of the primaries is $E_f=45$~GeV.}
\begin{ruledtabular}
\begin{tabular}{lr}
$p$ & $E_h$ \\
    & $({\rm Gy}/p)$ \\
\tableline
$e^\pm$    &  $(2.22\pm0.10)\times10^{-12}$ \\
$\pi^\pm$  &  $(5.03\pm0.63)\times10^{-14}$ \\
$K^\pm$    &  $(4.06\pm2.07)\times10^{-14}$ \\
$K_L$      &  $(3.62\pm1.03)\times10^{-14}$ \\
$K_S$      &  $(3.39\pm0.96)\times10^{-14}$
\end{tabular}
\end{ruledtabular}
\label{eh}
\end{table}

\end{document}